\documentstyle[11pt,paspconf,epsf]{article}

\begin{document}

\title{NGC\,2613, 3198, 6503, 7184: Case studies against `maximum' disks}

\author{B. Fuchs}
\affil{Astronomisches Rechen-Institut, 69120 Heidelberg, Germany}

\begin{abstract}
Decompositions of the rotation curves of NGC\,2613, 3198, 6503, and 7184 are
analyzed. For these galaxies the radial velocity dispersions of the stars
have been measured and their morphology is clearly discernible. If the
parameters of the decompositions are chosen according to the `maximum' disk
hypothesis, the Toomre $Q$ stability parameter is sytematically less than one
and the multiplicities of the spiral arms as expected from density wave theory
are inconsistent with the observed morphologies of the galaxies.
The apparent $Q\widetilde{<}1$ instability, in particular, is a strong
argument against the `maximum' disk hypothesis.
\end{abstract}


\section{Introduction}

`Maximum' disk versus submaximal disk decompositions of the rotation curves of
spiral galaxies have been discussed at great length in the literature (cf.~the
articles by Bosma and Sellwood in this volume). The aim of this paper is to
draw attention to the implications of such models of the rotation curves for
the internal dynamics of the disks.

The sample of galaxies studied here has been drawn from the list of Bottema
(1993) of galaxies with measured stellar velocity dispersions. The criteria
were: (a) the rotation curve of each galaxy, preferentially in HI, is observed,
(b) each galaxy is so inclined that the planar velocity dispersions are
measured, but (c) that its morphology is still discernible.

\section{Decomposition of the rotation curves and diagnostic tools}

The rotation curve of each galaxy is fitted by the superposition of
contributions due to the stellar and gaseous disks, both modelled by thin
exponential disks, the bulge (NGC\,2613 only), modelled by a
softened $r^{-3.5}$ density law, and the dark halo,
modelled by a quasi-isothermal sphere,
\begin{equation}
v_{\rm c}^2(R) = v_{\rm c,d}^2(R) + v_{\rm c,g}^2(R) + v_{\rm c,b}^2(R) +
v_{\rm c,h}^2(R)\,.
\end{equation}
The radial scale lengths of the disks, $h$, and core radii of the bulges,
$r_{\rm c,b}$, as well as the bulge to disk ratios have been adopted from
published photometry of the galaxies (cf.~Bottema (1993), Broeils (1992) and
references therein). Only in the cases of NGC\,3198 and 6503
HI data were available, which allowed the determination of the $v_{\rm c,g}$
contribution in Eq.~(1). No quantitative photometry of the bulge of NGC\,7184
is available.

   \begin{figure}[htbp]
   \begin{center}
   \epsfxsize=11.5cm
   \leavevmode
   \epsffile{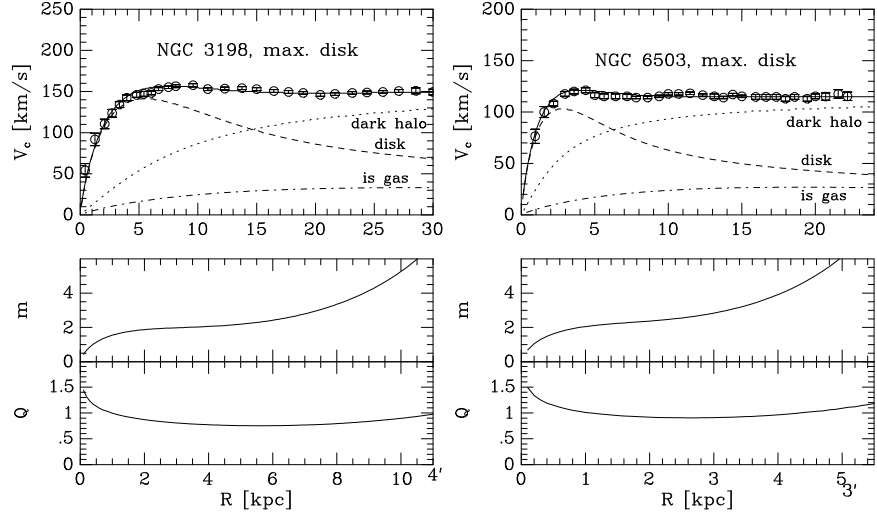}
   \end{center}
   \vspace*{-0.5cm}
\caption{`Maximum' disk decompositions of the rotation curves. The expected
   multiplicities of spiral arms and the stability parameters are shown in
   the lower panels.} \label{fig-1a}
   \end{figure}
The diagnostic tools, which I use to analyze the rotation curve models, are the
Toomre stability parameter of the disks and, following Athanassoula et
al.~(1987), the predicted multiplicity of the spiral structures. The Toomre
stability parameter is given by
\begin{equation}
Q = \frac{\kappa \sigma_{\rm U}}{3.36 G \Sigma_{\rm d}}.
\end{equation}
In Eq.~(2) $\kappa$ denotes the epicyclic frequency, which can be directly
derived from the rotation curve, $\sigma_{\rm U}$ the -- measured -- radial
velocity dispersion of the stars, $G$ the constant of gravitation, and
$\Sigma_{\rm d}$ the surface density of the disk, which follows from the fits
to the rotation curves. The stability parameter must lie in the range 1 $< Q
<$ 2, in order to prevent Jeans instability of the disk, on one hand, and
to allow the disks to develop spiral structures, on the other hand. All the
galaxies studied here are not grand-design spirals. In these galaxies the
spiral structures are almost certainly due to `swing amplification' of
perturbations of the disks (Toomre 1981). This mechanism is most effective, if
the circumferential wave length of the density waves is
\begin{equation}
\lambda = X\left(\frac{d v_{\rm c}(R)}{d R}\right) \lambda_{\rm crit} =
X\left(\frac{d v_{\rm c}(R)}{d R}\right) \frac{4 \pi^2 G \Sigma_{\rm
d}}{\kappa^2}\,.
\end{equation}
The value of the $X$ parameter is about 2 in the case of a flat rotation curve,
but less in the rising parts of the rotation curve (Athanassoula et al.~1987).
I apply
in Eq.~(3) a relation for $X(\frac{d v_{\rm c}(R)}{d R})$ found by analyzing
the stellardynamical equivalent of the Goldreich \& Lynden-Bell sheet (Fuchs
1991). The expected number of spiral arms is given by $m = 2 \pi R / \lambda$.
Eq.~(3) is derived from local density wave theory. Recent
alternative approaches based on global analyses of non-axisymmetric
perturbations of galactic disks are described by Haga \& Iye (1994),
Evans \& Read (1998a, b), and Pichon \& Cannon (1998).

Decompositions of the rotation curves of the galaxies, which maximise the disk
contribution in Eq.~(1), are shown in Figs.~1a and b together with
the resulting stability parameters and expected multiplicities of spiral arms.
As can be seen from Figs.~1 the $Q$ parameters are systematically close to or
even less than one. That is impossible in real galactic disks. As is well known
since the classical paper by Sellwood \& Carlberg (1984), the disks would
evolve fiercely under such conditions and heat up dynamically on short time
scales. If the model of Sellwood \& Carlberg is scaled to the dimensions of
NGC\,6503, the numerical simulations indicate that the disk would heat up
within a Gyr from $Q$ = 1 to 2.2 and any spiral structure would be
suppressed. The
amount of young stars on low velocity dispersion orbits, which would have to be
added to the disk in order to cool it dynamically back to $Q$ = 1, can be
estimated from Eq.~(2). In NGC\,6503 a star formation rate of 40
$M_\odot/pc^2/Gyr$ would be needed, while actually a star formation rate of 1.5
$M_\odot/pc^2/Gyr$, as deduced from the H$_\alpha$ flux (Kennicutt et
al.~1994), is observed. Thus `maximum' disks seem to be unrealistic under this
aspect.
   \begin{figure}[htbp]
   \begin{center}
   \epsfxsize=11.5cm
   \leavevmode
   \epsffile{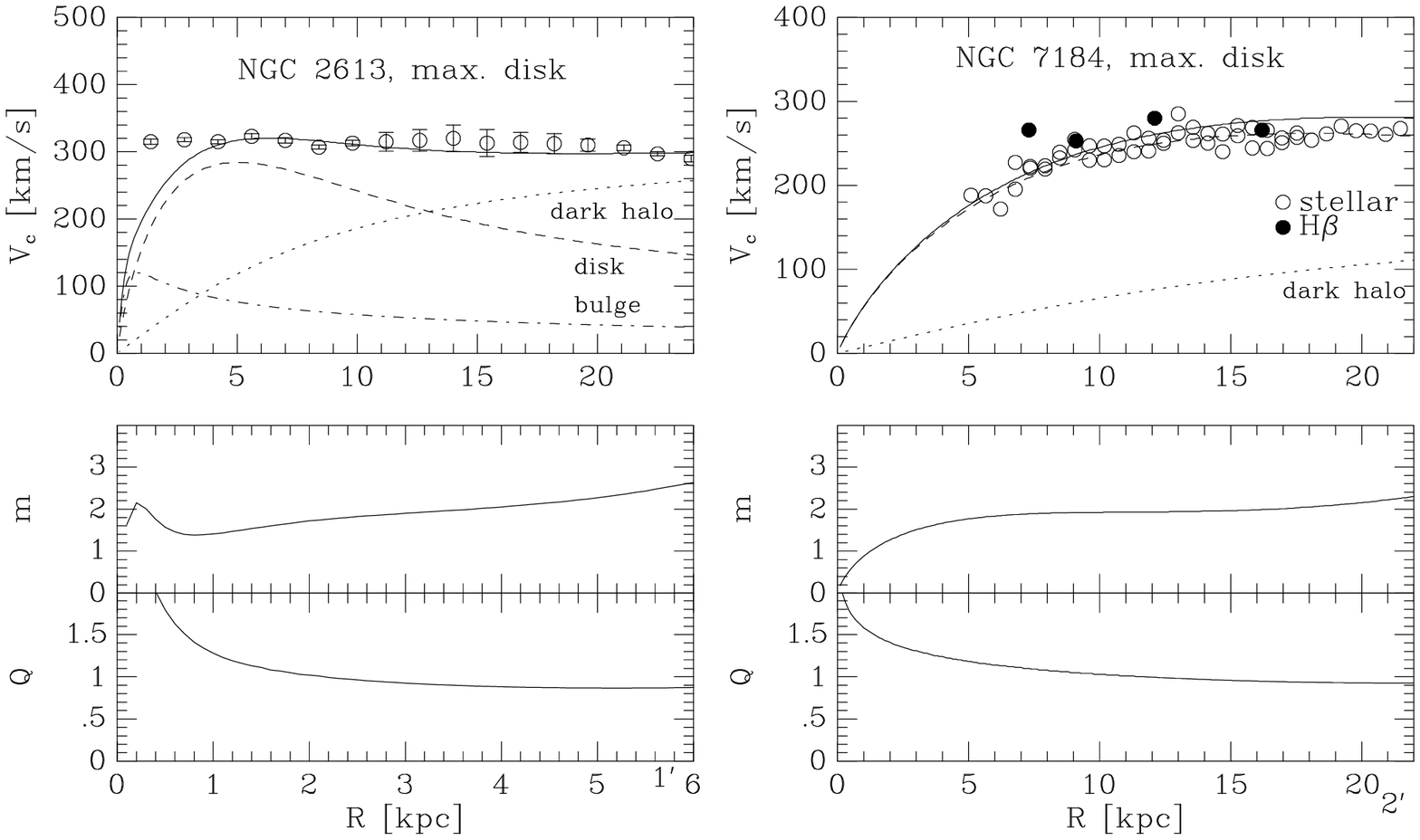}
   \end{center}
   \label{fig-1b}
   \end{figure}

Furthermore, according to the `maximum' disk models the galaxies should be
two-armed spirals. This is in agreement with Athanassoula et al.~(1987) and
Haga (1998, private communication), who concluded that in NGC\,3198 and 6503
$m$ is at least two. However, as can be seen on images of the galaxies
(cf.~The Carnegie Atlas of Galaxies), all the galaxies discussed here have
a multi-armed, irregular morphological appearrance.

Both deficiencies can be remedied simultaneously, if submaximal disks are
assumed. This is illustrated in Fig.~2 for NGC\,3198, where the mass-to-light
ratio of the disk has been reduced from $M/L_{\rm B}$ = 3.5
to  2.2 $M_\odot/L_{{\rm B},\odot}$. Within the
optical radius the dark halo contributes twice the mass of the disk and its
core radius is of the order of the radial scale length of the disk. As can be
seen from Fig.~2, the $Q$ parameter lies in a more realistic range and the
predicted multiplicity of spiral arms fits better to the observed morphology
of the galaxy than in the `maximum' disk model.

I am grateful to E.~Athanssoula, A.~Bosma, and J.~Kormendy for
valuable hints and discussions.
   \begin{figure}[htbp]
   \begin{center}
   \epsfxsize=11.5cm
   \leavevmode
   \epsffile{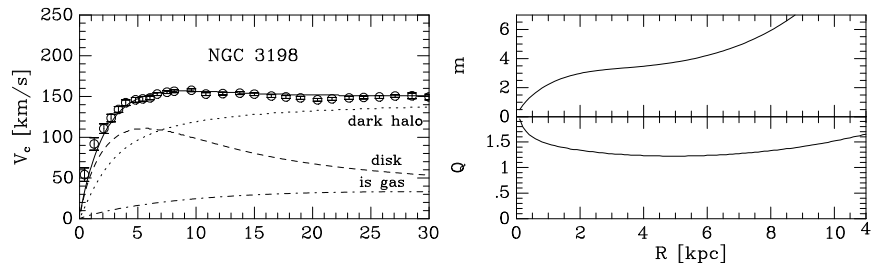}
   \end{center}
   \vspace*{-0.5cm}
\caption{Decomposition of the rotation curve with a submaximal disk.}
   \label{fig-2}
   \end{figure}
\small{
\begin{question}{Kormendy}
I'm worried about the interpretation of velocity dispersion measurements in
late-type disks that contain young stars. There is a danger that stars that
dominate the spectra are relatively bright, young, and low in velocity
dispersion and that the stars that dominate the disk mass are older, fainter
and hotter. Our Galactic disk shows just such an effect. It would imply that
measured dispersion values may be too small to correctly represent the disk
mass in calculations of $Q$.
\end{question}
\begin{answer}{Fuchs}
Yes, one has to keep this in mind. For the Galactic disk the effect can be
estimated using data from the solar neighbourhood. A detailed analysis
shows that the luminosity-weighted,
scale-height corrected radial velocity dispersion of stars in the Galactic disk
is $\sigma_{\rm U}$ = 36 km/s, which has to be compared with 44 km/s of the old
disk stars. The weight of young stars is 25\% of the total weight. In Sc
galaxies, which are bluer than the Galaxy with an averaged $<B-V>$ of 0.66 mag,
this might be shifted even more towards young stars. On the other hand, Sc
galaxies are more gas rich, which has a destabilizing effect. Taken all
together, the $Q$ argument seems to be quite robust.
\end{answer}
%\begin{question}{Gerhard}
%Radial scale lenghts of disks measured in the infrared tend to be smaller than
%optical scale lenghts, but may be more representative for the overall mass
%distributions in the disks. In what way would the stability results for the
%disks you studied be affected if smaller radial scale lengths are used?
%\end{question}
%\begin{answer}{Fuchs}
%If I use smaller scale lengths, I have to adjust the halo parameters in the
%fits to the rotation curves. The net effect is that the `maximum' disk
%models become more dark matter dominated. The outer parts of the disks are
%more stable, while the inner parts remain still too unstable.
%\end{answer}
}
\footnotesize{

}

\end{document}